# Thermodynamic and Quantum Thermodynamic Analyses of Brownian Movement

Elias P. Gyftopoulos*
Massachusetts Institute of Technology
Department of Nuclear Engineering and Mechanical Engineering
Cambridge, MA  02139

Thermodynamic and quantum thermodynamic analyses of Brownian movement of a solvent and a colloid passing through neutral thermodynamic equilibrium states only.  It is shown that Brownian motors and *E. coli* do not represent Brownian movement.

Brownian movement has a history of more than two centuries.  Studies that appeared in the scientific archival literature in the period until 1920 are listed by Einstein [1].  Articles that were written until 1970 are presented in Refs. [2] and [3].  Three-dimensional Brownian motors are discussed in [4], and Brownian movement of *E. coli* in [5].  All these studies and articles are based on a variety of *statistical interpretations* of thermodynamics, and on the conception of the molecular-kinetic theory of heat.  Maxwell [6] conceived his omniscient and omnipotent demon who can contradict a circularly postulated second law of thermodynamics, and concludes … "In dealing with masses of matter, while we do not perceive the individual molecules, we are compelled to adopt what I have described as the statistical method of calculation, and to abandon the strict dynamical method in which we follow every molecule by the calculus".

Boltzmann interpreted the entropy $S$ as a measure of disorder, and specified the expression $S = \mathrm{k}\log\Omega$.  The idea of disorder has been adopted by many scientists [7-9].

Statistical theories of thermodynamics yield many correct and practical numerical results only about thermodynamic equilibrium states [10, 11].  Over the past almost two centuries however, despite these successes, thousands of scientists and engineers [12] have expressed a dissatisfaction with the almost universal efforts to compel thermodynamics to conform to statistical explanations in the light of both many accurate, reproducible nonstatistical experiences and many theoretical inconsistencies that have been identified, and a desire for a better theory as proposed for the first time by Carnot [13].

In response to these concerns we have developed two expositions of thermodynamics, one without reference to quantum theory that applies to all systems (large and small), and to all states (thermodynamic and non-thermodynamic equilibrium [14], and the other quantum theoretic without statistical probabilities [15-17]. In the light of these expositions, we prove the following results that are relevant to Brownian movement.

*Email: dboomer@mit.edu

Systems in which Brownian movement is observed are in a neutral thermodynamic equilibrium state, and consist of two parts: (i) a liquid solvent, and (ii) a colloid composed of particles much larger than atoms but too small to be visible by an unaided eye, and dispersed but not dissolvable by the solvent.

Each of the two parts has the same temperature $T$, and the same pressure $p$. Moreover, we prove that if a constituent is not present in the solvent or the colloid, then its total potential $\mu$ is minus infinity [18]. So, for the i-th constituent of the solvent and the j-th constituent of the colloid we have the following inequalities

$$(\mu_{\text{solvent}})_i > (\mu_{\text{colloid}})_i = -\infty \tag{1}$$

$$(\mu_{\text{colloid}})_j > (\mu_{\text{solvent}})_j = -\infty \tag{2}$$

An implication of inequalities (1) and (2) is that the solvent and the colloid are in partial mutual stable equilibrium, that is, they satisfy the conditions of temperature and pressure equalities but not the conditions of total potential equalities. As a result, both the constituents of the solvent and the colloid exert infinitely large "driving forces" (total potential differences) on the pliable interface between the two parts, and try to interpenetrate each other as they would have done if the colloid were soluble by the solvent. However, such an interpretation is impossible, and the only effect is a continuous in time modification of the pliable shape of the interface, a modification that does not affect the entropy, the energy, the volume, and the amounts of constituents of either the solvent or the colloid and therefore the temperature, the pressure, and the total potentials of the composite of these two systems. Said differently, it is not the motions of the solvent and the colloid that cause the observed movements but the infinitely large differences in total potentials that change the shape of the interface and appear to the observer as motions.

It is clear that controllable three-dimensional motors [4] in symmetric potentials do not represent Brownian movement because whatever is observed is the result of interactions of a system with systems in its environment. No such interactions apply to a system consisting of a solvent and a colloid.

The resolution of the dilemmas and paradoxes that have preoccupied generations of physicists over more than a century in their attempts to rationalize the relation between mechanics and thermodynamics without statistics has been achieved as follows:

(i) The recognition that the quantum-mechanical density operators $\rho \geq \rho^2$ that are subject to the laws of physics (quantum-theoretic and thermodynamic) are those that can be represented by a *homogeneous ensemble*. In such an ensemble, every member is assigned the same $\rho$ as any other member, and experimentally (in contrast to algebraically) $\rho$ cannot be decomposed – is unambiguous or irreducible – into a statistical mixture of either projectors or density operators different from ρ [15]. The relevance and reality of unambiguous density operators has also been identified by Jauch [19], and had been observed by Schrödinger [20] who, however, did not pursue the consequences of his observation.

(ii) The recognition that the Schrödinger equation of motion is correct but incomplete. It is incomplete because it describes only evolutions in time that are unitary and therefore reversible. The same remarks apply to the von Neumann equation of motion for statistical density operators. But not all reversible evolutions in time are unitary and not all evolutions are reversible. In response to this recognition, Beretta et al [16, 17] conceived a nonlinear equation of motion for ρ that has as a limiting case the Schrödinger equation (zero entropy physics) and regularizes evolutions in time that are reversible and either unitary or nonunitary, and evolutions that are irreversible.

(iii) The determination of the analytical expression for entropy [21] which differs from each and every of the dozens of expressions that have been proposed in the literature but is the only one that satisfies nine criteria that have been established in the quantal and nonquantal expositions of thermodynamics. It is given by the relation

$$S = -k\text{Tr}[\rho \ln \rho]$$

where ρ is represented by a homogeneous ensemble and not a statistical average of projectors.

(iv) The interpretation of the entropy of quantum thermodynamics as a measure of the spatial shape of the constituents of the system in any state, stable equilibrium or not stable equilibrium [22, 23]. None of the statistical entropies have such characteristics.

(v) Whereas the entropies of statistical mechanics are thought to represent ultimate disorder if the system is in a stable equilibrium state, the entropy of the unified theory represents perfect order for such a state [24].

For the quantum thermodynamic analysis we consider the same system and state as for the thermodynamic analysis. While the solvent and the colloid pass through neutral stable equilibrium states, the energy eigenprojectors and eigenvalues of each subsystem change continuously so as to accommodate the continuous changes of the shapes of the constant volumes.

These continuous changes in time are impossible to evaluate both because of lack of knowledge of the precise changes of the shapes of the volumes, and the difficulty inherent in calculating eigenprojectors and eigenvalues in cases of complicated shapes of even very simple systems such as one particle in an odd looking, one dimensional potential well. The laws of physics, however, have no difficulty in continuously in time responding to the changing shapes of the liquid solvent and the colloid and determining the nonstatistical density operators for each pair of shapes at each instant in time.

An explicit analysis of the relations between occupation probabilities and energy eigenvalues has been performed by Çubukçu [25]. He considers the Hamiltonian operators $H_g$ for $g = s, c$, where s is the solvent and c the colloid, and finds the following results: Hamiltonian operator of the two systems $H = H_s \otimes I_c + I_s \otimes H_c$ where I is the identity operator. Energy, eigenfunctions and eigenvalues $H_s \psi_i = e_i \psi_i$, $H_c \varphi_j = \varepsilon_j \varphi_j$, $H(\psi_i \otimes \varphi_j) = (e_i + \varepsilon_j)(\psi_i \otimes \varphi_j)$, density operator of the system $\rho(t) = \rho_s(t) \otimes \rho_c(t)$, eigenvalues of the overall density operator $\rho(t)(\psi_i \otimes \varphi_j) = p_{ij}(\psi_i \otimes \varphi_j)$, relations between density operator eigenvalues and energy eigenvalues $\ln(p_{ij}/p_{kl}) = [(e_k - e_i) + (\varepsilon_l - \varepsilon_j)]/kT$ for all pairs {i, j} and {k, l}, where *T* is the constant temperature of the solvent and the colloid. From this analysis we see that as the eigenvalues and eigenfunctions of the solvent and the colloid change, the constituents of each of the two substances must be reallocated to the evolving energy eigenstates with evolving occupation probabilities, and the reallocation appears as Brownian movement.

A description of the behavior of *E. coli* is given in [5]. Though the description is accurate, our view is that it is not related in any way whatsoever to Brownian motion for the following reasons. In contrast to systems in which Brownian movement is observed, and which consist of a liquid solvent and a colloid, each maintaining its identity for all practical purposes for ever, *E. coli* have a totally different biography. Relevant statements that one finds in textbooks on molecular biology of the gene [26] and biochemistry [27] are as follows: "…the bacteria *E. coli* will grow in an aqueous solution containing just glucose and several inorganic ions" … "There is a lower limit, however, to the time necessary for a cell generation; no matter how favorable the growth conditions, an *E. coli* is unable to divide more than once every 20 minutes. … The average *E. coli* cell is rod shaped. … It grows by increasing in length followed by a fission process that generates two cells of equal length." In addition, *E. coli* is self propelled not as a result of infinite differences between total potentials such as exist between a liquid and a solvent and a colloid but because of flagella.

In view of all these facts, one must conclude that the time dependent changes of *E. coli* studied by molecular biologists, and biochemists do not represent Brownian motions.